\documentclass[conference,10pt]{IEEEtran}

\usepackage[T1]{fontenc}
\usepackage[utf8]{inputenc}
\usepackage[english]{babel}
\usepackage{microtype}
\usepackage{graphicx}
\usepackage{booktabs}
\usepackage{xcolor}
\usepackage{float}
\usepackage{hyperref}
\usepackage{orcidlink}
\usepackage{balance}
\usepackage{csquotes}
\usepackage{tabularx}
\usepackage{fancyhdr}

\newcolumntype{Y}[1]{>{\hsize=#1\hsize\raggedright\arraybackslash}X}

\usepackage[
  backend=biber,
  style=ieee,
  sorting=none,
  doi=true,
  url=true,
  isbn=false
]{biblatex}
\newif\ifanon
\anonfalse

\title{From a Last-Week Add-On to First-Day Practice:\\
Three Years of Integrating MPI Performance Analysis into HPC Education with EduMPI Suite}

\ifanon
  \author{\IEEEauthorblockN{Anonymous Authors}}
  \hypersetup{
    pdftitle={From a Last-Week Add-On to First-Day Practice: Three Years of Integrating MPI Performance Analysis into HPC Education with EduMPI Suite},
    pdfsubject={EduHPC-26 Experience/Replication Paper},
    pdfauthor={Anonymous Authors},
    pdfkeywords={HPC Education, MPI, Performance Analysis, Visualization, EduMPI, Experience Report}
  }
\else
  \author{
    \IEEEauthorblockN{
      Anna-Lena Roth\,\orcidlink{0000-0002-6463-3486}\IEEEauthorrefmark{1},
      Jonas Posner\,\orcidlink{0000-0002-6491-1626}\IEEEauthorrefmark{1}
    }
    \IEEEauthorblockA{
      \IEEEauthorrefmark{1}Fulda University of Applied Sciences, Germany\\
      \{anna-lena.roth, jonas.posner\}@cs.hs-fulda.de
    }
  }
  \hypersetup{
    pdftitle={From a Last-Week Add-On to First-Day Practice: Three Years of Integrating MPI Performance Analysis into HPC Education with EduMPI Suite},
    pdfsubject={EduHPC-26 Experience/Replication Paper},
    pdfauthor={Anna-Lena Roth, Jonas Posner},
    pdfkeywords={HPC Education, MPI, Performance Analysis, Visualization, EduMPI, Experience Report}
  }
\fi

\AtEveryBibitem{
  \clearfield{eprint}
  \clearfield{eprinttype}
  \clearfield{archiveprefix}
  \clearfield{urldate}
}

\newfloat{listing}{t}{lol}
\floatname{listing}{Listing}

\newcommand{\acceptedManuscriptNotice}{%
    \textit{Author accepted manuscript. Accepted for publication in the SC26 Workshops.}\par
    \smallskip
    \copyright\ 2026 IEEE. Personal use of this material is permitted.
    Permission from IEEE must be obtained for all other uses, in any current
    or future media, including reprinting/republishing this material for
    advertising or promotional purposes, creating new collective works,
    for resale or redistribution to servers or lists, or reuse of any
    copyrighted component of this work in other works.
}

\fancypagestyle{firstpagecopyright}{%
  \fancyhf{}%
  \renewcommand{\headrulewidth}{0pt}%
  \renewcommand{\footrulewidth}{0pt}%
  \fancyfoot[C]{\raisebox{18pt}[0pt][0pt]{%
    \parbox[t]{\textwidth}{\footnotesize\acceptedManuscriptNotice}}}%
}

\begin{document}

\maketitle

\thispagestyle{firstpagecopyright}

\begin{abstract}
Performance analysis is essential for scalable MPI programs, yet tool and workflow complexity often relegates it to the end of Parallel Programming courses.
EduMPI Suite lowers this entry barrier by automating cluster execution, measurement, and near-real-time analysis while visualizing MPI communication from the first program.
This paper presents a three-year experience report on its integration into HPC education.
We trace the transition of performance analysis from a last-week add-on to a recurring element of MPI development, using unpublished historical survey evidence and individual 2025 questionnaire data.
Earlier cohorts perceived performance analysis as unnecessary, difficult, or unintuitive.
In the final study, all students recommended introducing EduMPI Suite early in the course.
Of 33 students, 29--33 rated it higher than conventional tools for understanding program behavior, interpreting communication, identifying bottlenecks, obtaining visual feedback, and navigating the tool.
These findings support embedding performance analysis throughout MPI education before transitioning learners to professional environments.
\end{abstract}


\section{Introduction}
The Message Passing Interface (MPI) is the dominant programming model for distributed-memory High Performance Computing (HPC) systems: developers decompose work across processes and exchange data explicitly through point-to-point, one-sided, and collective communication.
Efficient MPI programming therefore requires more than syntactically correct code.
The same program can behave differently depending on process placement, node topology, memory hierarchy, network characteristics, and workload manager decisions.
This performance-portability problem is part of everyday parallel programming: experienced developers analyze applications on the cluster where they will run and use evidence from that system to locate waiting states, load imbalance, and communication volume.
These competencies therefore need to be acquired during programming education.
Current HPC curricula accordingly recommend practical performance analysis and visualization as part of parallel-programming instruction~\cite{bouvry2025,prasad2020,raj2020,roth2023survey}.

Professional analysis workflows, however, are difficult to introduce when beginners first write MPI code.
Score-P~\cite{knuepfer2012} is an open, widely used instrumentation and measurement infrastructure.
The resulting traces and profiles can be explored with established tools such as the commercial Vampir~\cite{KnuepferBrunst2008} and open tools including CUBE~\cite{CubeV4} and TAU~\cite{ShendeMalony2006}.
These tools are powerful, but effective use requires cluster operation, instrumentation, measurement, and tool-specific interpretation: students must know which performance metrics to inspect, where performance problems arise in code, and which view can confirm a hypothesis, while also handling SSH, Linux commands, file transfer, compilation, job scripts, and workload managers such as SLURM. 
The usual \emph{post-mortem} workflow adds further friction because data become visible only after instrumentation, a measurement run, and result-file generation.
In education, this combined complexity often pushes performance analysis toward the course end, where it can appear as a specialized topic rather than an integral part of MPI development.
Our master's-level Parallel Programming course at Fulda University of Applied Sciences, Germany, followed that conventional response until 2023: teach MPI first and introduce performance analysis in the final exercise.

EduMPI Suite was developed to invert this sequence.
It automates cluster use: students start MPI programs from EduMPI GUI on their local machines and execute them on the cluster without first mastering SSH, job scripts, or workload-manager commands.
Performance analysis is triggered automatically, and performance data are shown in near-real-time in the GUI.
Students are therefore confronted with concrete MPI communication behavior while they are executing the program on a cluster, not only after a separate post-mortem workflow.
EduMPI Suite presents this behavior through complementary 2D and 3D views that connect ranks, messages, time intervals, and physical placement.
As a novel feature, it also opens the black box of collective communication by visualizing the internal point-to-point exchanges selected by the MPI library, which remain largely hidden in professional tools~\cite{roth2025collectives}.
The goal is not to replace professional performance analysis, but to make it available from the first MPI program onward.

This paper reports three years of integrating performance analysis into our Parallel Programming course and developing EduMPI Suite through classroom use and usability studies. 
We ask: \emph{How did moving from last-week professional-tool exercises to first-day EduMPI-supported practice affect students' perception and use of MPI performance analysis?} We contribute (1) previously unpublished survey evidence from two historical end-of-course TAU cohorts; (2) a synthesis of the design decisions that shaped EduMPI Suite across the three-year development trajectory; and (3) a previously unpublished individual item-level comparison from a 2025 usability study in which students evaluated EduMPI Suite against TAU or CUBE.
Together, they show how EduMPI Suite strengthens performance-analysis understanding while bridging to professional environments.

\section{EduMPI Suite Background}
\subsection{Architecture and workflow}
EduMPI Suite consists of three core components~\cite{roth2025system,roth2026architecture}. 
\emph{EduMPI} is a fork of Open MPI that preserves standard MPI functionality while adding measurement of communication events. 
\emph{EduStore} is a time-series database that receives these events from running applications and supports low-latency queries. 
\emph{EduMPI GUI} is the cross-platform desktop interface through which students execute programs and inspect the collected data. 
It is available for common student operating systems, including Windows, Linux, and macOS.

The resulting workflow hides the operational steps that normally precede performance analysis. 
Students select source files and process counts in the GUI.
EduMPI Suite transfers the files, generates the SLURM job, compiles and launches the application, and monitors its state on the cluster. 
While the program is running, EduMPI streams MPI event data to EduStore and EduMPI GUI queries them for near-real-time visualization. 
For responsiveness in classroom use, EduMPI GUI primarily shows data aggregated per process and second.
Students can select intervals, abort runs, review experiments, and automatically generate Score-P profiles for professional visualization tools~\cite{roth2024delfi,roth2025system}.
Evaluated runtime overhead remains below 3.8\%, and the architecture has supported concurrent classroom use~\cite{roth2026architecture}.

\subsection{MPI communication as the central representation}
EduMPI Suite was designed around MPI communication rather than around general-purpose profiling tables. 
Its central representation connects source-level MPI behavior to physical cluster structure. 
A 3D overview places processes inside nodes and visualizes communication partners by connecting them with arrows, while 2D views expose detailed data volumes.
Runtime options represent send/receive ratios across processes as well as waiting states.

The most distinctive feature is the visualization of collective communication. 
While professional tools display \texttt{MPI\_Bcast} as a collective event, EduMPI Suite exposes MPI's underlying point-to-point messages~\cite{roth2025collectives}.
This makes hidden implementation decisions observable: students can compare a manually implemented broadcast with \texttt{MPI\_Bcast}, identify bottlenecks, and explain differences in communication structure and scalability.

\subsection{Prior evaluations}
EduMPI Suite was refined through several user and usability tests.
In winter semester (WiSe) 2023/24, a preliminary usability study with a conventional Score-P/Scalasca/CUBE workflow showed that EduMPI Suite reduced the median time to the first visualization from about 26 minutes to about 5 minutes and reduced moderator interventions during performance instrumentation and measurement from a median of eight to one~\cite{roth2024delfi}.
This motivated the focus on automated cluster execution and immediate near-real-time feedback.

Classroom tests in WiSe 2024/25 further examined collective communication.
Among 30 students, 29 recognized \texttt{MPI\_Bcast} as more efficient than an iterative point-to-point broadcast, 28 identified rank~0 as the bottleneck of the iterative point-to-point broadcast, and students identified message-size-dependent algorithm changes within \texttt{MPI\_Bcast}~\cite{roth2025collectives}.

A comprehensive comparative usability study in summer semester (SuSe) 2025 then evaluated the mature version of EduMPI Suite against TAU or CUBE with 33 master's students. 
Students solved 96\% of analysis tasks with EduMPI Suite compared with 60\% using the professional tools, required fewer interventions, and all 33 recommended EduMPI Suite for educational use~\cite{roth2026usability}. 
The counterbalanced design also indicated transfer: groups using EduMPI Suite first later performed better with TAU/CUBE than groups starting with the professional tools.
These studies establish the tool's usability trajectory.

This paper adds the longitudinal classroom interpretation, previously unpublished historical baseline evidence, and the previously unpublished individual item-level comparison from the SuSe 2025 study.

\section{From Add-On to First-Day Practice}
\subsection{The baseline workflow}
In SuSe 2022 and WiSe 2022/23, performance analysis appeared at the end of the course, after students had already completed most MPI programming tasks~\cite{roth2026historicaldata}.
Students installed TAU locally, instrumented an MPI application with Score-P, submitted cluster jobs, transferred result files, and analyzed them with TAU after an in-class demonstration.
This baseline required novices to solve operational, representational, and conceptual problems simultaneously.
Visualization was available only after a successful post-mortem pipeline, while learners still had to know which metric or view to inspect.

\begin{figure*}[t]
\centering
\includegraphics[width=0.75\textwidth]{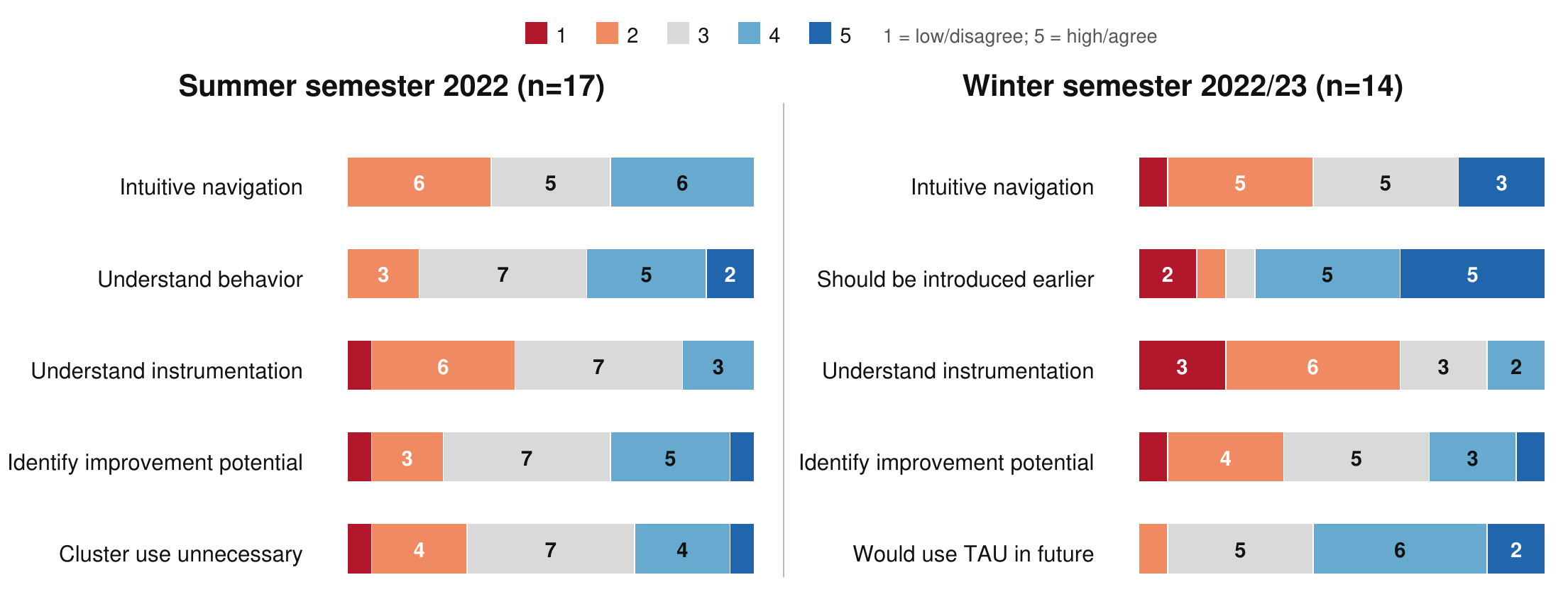}
\caption{Likert distributions from two end-of-course TAU surveys. Numbers inside segments are response counts; unlabeled segments contain one response.}
\label{fig:historical}
\vspace*{-0.25cm}
\end{figure*}

In SuSe 2022, only 6 of 17 respondents rated TAU's navigation intuitive, 7 agreed that TAU helped them understand program behavior, and 6 independently identified improvement potential (Fig.~\ref{fig:historical}). 
Just 3 agreed that they understood code-instrumentation principles, while 5 considered local execution sufficient and cluster use unnecessary. 
Perceived relevance did not reliably translate into usable analysis or cluster practice.

Only 3 of 14 rated TAU's navigation intuitive, 4 independently identified improvement potential of an MPI program, and 2 agreed that they understood the instrumentation and measurement workflow.
At the same time, 10 wanted earlier analysis and 8 intended to reuse TAU. 
Across the three shared dimensions, positive ratings remained a minority in both cohorts: navigation (6/17 and 3/14), instrumentation (3/17 and 2/14), and independent identification of improvement potential (6/17 and 4/14). 
Late placement generated interest but left insufficient time to build independent competence. 
In the WiSe 2023/24 conventional-tool study, six students described analysis as unnecessary or too difficult and suggested it, at most, as a bonus or final topic~\cite{roth2024delfi}.

\subsection{Evolution of the intervention}
EduMPI Suite was developed in response to the limitations of the conventional sequence used in SuSe 2022 and WiSe 2022/23, in which students first learned MPI programming and encountered TAU only in the final exercise.
These limitations motivated a first major design decision: performance analysis should become an integral part of learning MPI rather than remain a final add-on.
Accordingly, EduMPI Suite was designed to reduce operational barriers such as SSH access, job scripts, and workload managers, while providing immediate, near-real-time feedback from the first MPI program onward.

The next iterations addressed barriers exposed by classroom use and usability studies.
The WiSe 2023/24 Score-P/Scalasca/CUBE study showed that conventional workflows consumed substantial time and instructor support before a first visualization~\cite{roth2024delfi}.
EduMPI Suite therefore automated cluster access, job generation, measurement, and near-real-time visualization. 
Once this made early use possible, the problem shifted from access to explanation: students needed views that connected source-level MPI calls to ranks, nodes, cores, communication volume, and phases of execution. 
This motivated the timeline, time selection, topology-aware 2D/3D views, communication arrows, and per-rank data later described as core EduMPI Suite features~\cite{roth2025system}.

The user tests focusing on collective communication extended this idea from visible user-level messages to MPI-internal behavior.
Because professional tools usually present collectives as library calls, students cannot easily see why \texttt{MPI\_Bcast} differs from a manually implemented broadcast or why message size changes the algorithm selected. 
EduMPI Suite was therefore extended to expose the internal point-to-point exchanges used by the MPI implementation behind collectives, making its selected communication pattern visible; the teaching tasks were redesigned around explaining these algorithmic patterns~\cite{roth2025collectives}.

Finally, the comprehensive 2025 usability study showed both the strength and the boundary of this scaffold: EduMPI Suite supported local diagnosis and mental-model formation, while CUBE and TAU remained valuable for aggregated post-mortem performance analysis~\cite{roth2026usability}. Table~\ref{tab:evolution} summarizes the resulting evaluation-to-design loop.

\begin{table*}[t]
  \vspace*{-0.5cm}
  \caption{Evolution of the classroom intervention and evidence base.}
  \label{tab:evolution}
  \centering
  \scriptsize
  \setlength{\tabcolsep}{3.5pt}
  \begin{tabularx}{\textwidth}{@{}p{1.55cm}Y{0.813}Y{0.973}Y{1.214}@{}}
      \toprule
      Phase & Classroom intervention & Observed barrier or opportunity & Resulting design/teaching decision \\
      \midrule
      SuSe 2022 / WiSe 2022/23 & TAU in the final exercise & Guided TAU use was possible and partly valued, but analysis remained a late specialist topic and students rarely identified improvement potential without support & Move from one final TAU session to recurring performance-analysis tasks throughout the course; remove operational barriers \\
      WiSe 2023/24 & Preliminary usability study: Score-P/Scalasca/CUBE vs. EduMPI Suite & Conventional workflow required much more time and command-level help before first visualization & Automate cluster login, job generation, measurement, and visualization in the GUI; support repeated runs~\cite{roth2024delfi} \\
      SuSe 2024 & EduMPI Suite during regular MPI exercises & Early execution worked, but explanations needed clearer links among code, ranks, nodes/cores, and phases & Add timeline/time selection and topology-aware 2D/3D views with communication arrows and per-rank data~\cite{roth2025system} \\
      WiSe 2024/25 & User studies on collective communication & Students needed evidence for why collectives outperform hand-written loops and how MPI changes algorithms & Expose internal collective point-to-point patterns and design tasks around explaining algorithm changes~\cite{roth2025collectives} \\
      SuSe 2025 & Comprehensive, counterbalanced EduMPI Suite--TAU/CUBE usability study & EduMPI Suite supported local diagnosis; professional tools remained useful for aggregation and detailed post-run views & Position EduMPI Suite as scaffold: start with visual hypotheses, then export/transfer to CUBE or TAU and add summaries~\cite{roth2026usability} \\
      \bottomrule
  \end{tabularx}
\end{table*}

The revised course uses EduMPI Suite from the first MPI program onward so that students learn to identify bottlenecks, waiting states, and late senders and relate them to their code and MPI communication.
It lowers the entry barrier by deferring SSH, SLURM, manual execution, and professional measurement workflows until later in the course, when students transition to CUBE or TAU for post-mortem analysis.

\section{Results After Three Years}
\subsection{From group-level findings to classroom experience}
After three years, the most important outcome is that performance analysis moved from a final exercise to a repeated way of reasoning about MPI programs. 
In the published comprehensive usability study, students solved 96\% of analysis tasks with EduMPI Suite compared with 60\% using TAU or CUBE; interventions were needed in 29\% versus 55\% of tasks; and EduMPI Suite was usually faster for locating processes, communication patterns, or bottlenecks~\cite{roth2026usability}. 
All 33 participants recommended EduMPI Suite for educational use, compared with one unqualified recommendation for TAU or CUBE.
The crossover design further showed a transfer effect: groups that used EduMPI Suite first solved more subsequent TAU/CUBE tasks correctly and needed fewer interventions with those professional tools.

EduMPI Suite was strongest when students connected a visible symptom to an MPI mechanism, such as waiting ranks, centralized communication, process placement, or collective algorithms. 
Its main weakness was showing summarized statistics about the whole program: because EduMPI Suite is optimized for near-real-time, time-segmented feedback, CUBE and TAU remain stronger for post-mortem profiles. 
We therefore treat EduMPI Suite as the first scaffold for forming hypotheses, followed by professional tools for aggregated post-mortem analysis.

\subsection{Previously unpublished individual comparison}
The published comprehensive usability study reported group-level task success, interventions, processing time, recommendations, and transfer observations~\cite{roth2026usability}. 
It did not present the individual post-test questionnaire data~\cite{roth2025data}. 
For this experience report, we use these individual ratings to show whether the favorable experience was concentrated in a few groups or visible across students.
The 33 participants worked in dyads or triads in a counterbalanced design: six groups used EduMPI Suite before TAU or CUBE, and six groups used the reverse order; nine groups selected CUBE and three selected TAU.
This resulted in 25 individual comparisons between EduMPI Suite and CUBE and eight between EduMPI Suite and TAU.

\begin{figure*}[t]
\centering
\includegraphics[width=0.75\textwidth]{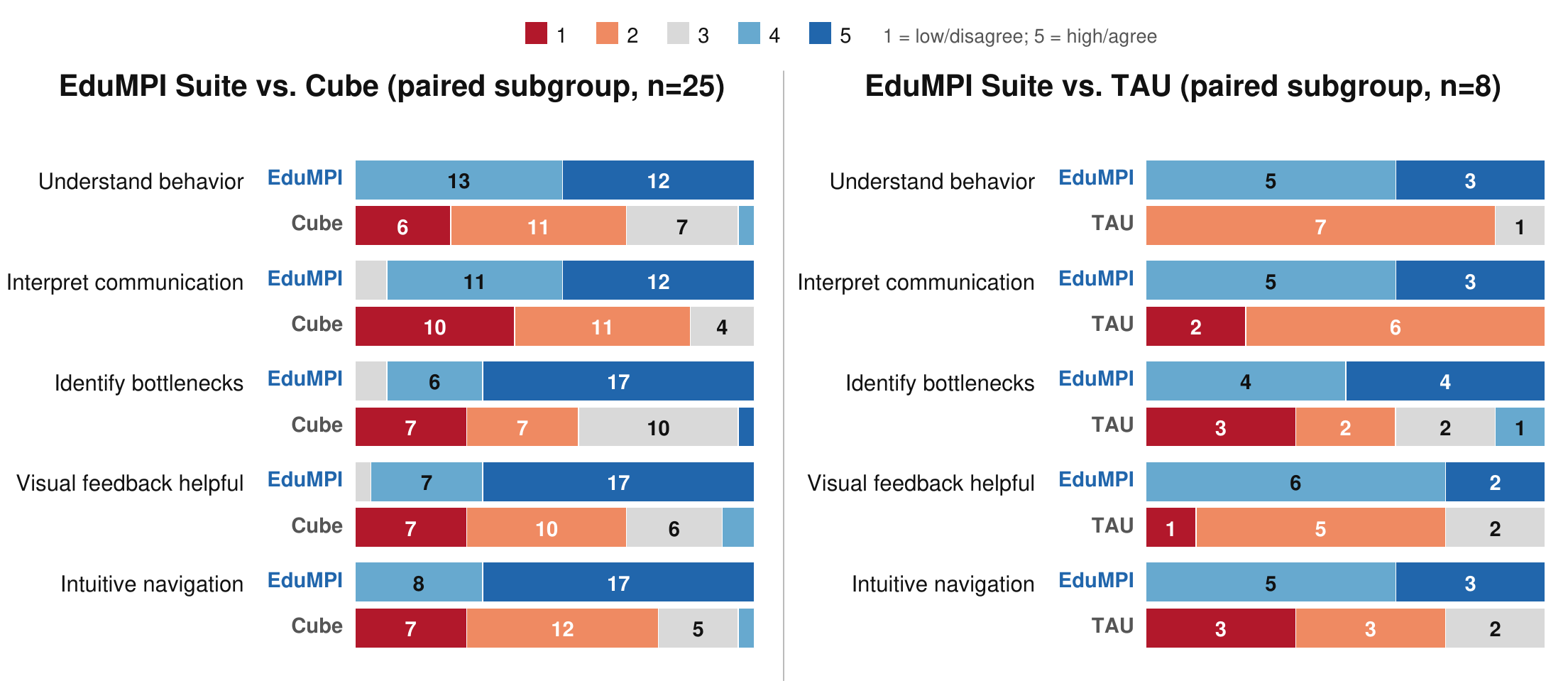}
\caption{Previously unpublished individual paired Likert distributions from the SuSe 2025 evaluation, separated by the professional tool selected by each group. Each participant rated EduMPI Suite and CUBE or EduMPI Suite and TAU on the same five items. Numbers inside segments are response counts; unlabeled narrow segments contain one or two responses.}
\label{fig:current}
\vspace*{-0.25cm}
\end{figure*}

The distributions in Fig.~\ref{fig:current} show that the result is not driven by one professional tool or by a small number of enthusiastic groups. 
In the CUBE subgroup, all 25 participants rated EduMPI Suite 4 or 5 for understanding program behavior and tool navigation, whereas 24 rated CUBE between 1 and 3 on each item. 
In the TAU subgroup, all eight rated EduMPI Suite 4 or 5 on every dimension; TAU received only one rating above 3. 
The paired direction is equally consistent: every participant rated EduMPI Suite higher for understanding program behavior, interpreting communication, and tool navigation. 
For bottleneck identification, 29 rated EduMPI Suite higher and four tied.
For visual feedback, only one participant rated the professional tool higher. 
These individual data strengthen the experience-report claim: the shift was not merely that groups completed more tasks, but that students individually perceived EduMPI Suite as the more helpful representation for the core activities of beginner-level MPI performance analysis.

\section{What Changed Compared with the Baseline}
In the late TAU baseline, positive ratings remained a minority for intuitive navigation (6/17 in 2022 and 3/14 in 2022/23), understanding instrumentation (3/17 and 2/14), and independently identifying improvement potential (6/17 and 4/14). 
In 2025, all 33 participants rated EduMPI Suite positively for understanding program behavior and tool navigation, and every participant rated it higher than TAU/CUBE for understanding program behavior, interpreting communication, and tool navigation. 
The comparison is not a controlled cohort study because wording, tasks, and timing changed, but the direction is consistent: the activity that had previously remained late and support-dependent became understandable enough to be used repeatedly during MPI development.

The second change concerns support needs and transfer. 
Beyond students' preference for EduMPI Suite, the higher task success, lower support needs, and subsequent TAU/CUBE results reported above support the teaching pattern that emerged from the three-year trajectory: first make evidence immediate, then build a visual communication model, reveal collective algorithms, and finally move to professional tools once students can formulate performance questions.

The instructor's role changes accordingly.
Because students need less procedural help with cluster access, measurement, and tool operation, the instructor can focus on conceptual questions: Which rank waits, how does doubling ranks change the pattern, and why does MPI select different broadcast algorithms for different message sizes?
Screenshots and time selections become evidence in explanations, making performance analysis assessable as reasoning rather than tool operation.

In practice, this changed the course culture. 
Performance analysis became a regular part of MPI programming: students increasingly asked not only whether a program was correct, but how well it behaved and where performance could be improved. 
Several students used screenshots from performance-analysis tools, especially EduMPI Suite, in final presentations to explain their programs and demonstrate observed communication behavior. 
The near-real-time visualization supported more independent cluster use: when students discovered a problem early, they could abort a run and free limited educational cluster resources instead of waiting for post-mortem results.

The approach also has limitations. 
While EduMPI Suite helps students relate visible symptoms to MPI mechanisms, its focus on near-real-time, time-segmented feedback makes aggregated profile overviews less accessible.
Questions such as which MPI call dominates the total runtime still motivate a handover to CUBE or TAU. 
Automation also did not make every representation intuitive: some students still criticized 3D navigation, ambiguous colors, and missing labels.
Finally, EduMPI Suite requires deployment effort: EduStore, cluster integration, and ongoing maintenance.
The architecture has been reliable with up to 24 simultaneous classroom users and low overhead~\cite{roth2026architecture}, but the evidence remains a single-institution experience with small historical cohorts and a 2025 study shaped by group work, prior EduMPI Suite exposure, and the selected comparison tools.\looseness=-1

For adoption elsewhere, the main lesson is to reproduce this sequence rather than treat EduMPI Suite as a replacement for professional tools.
Institutions must plan for this infrastructure and its maintenance, while retaining CUBE or TAU for aggregate post-mortem questions.
Less cluster-experienced populations may benefit especially from separating operational setup from performance reasoning; whether the same scaffold benefits other institutions or student populations, however, requires evaluation beyond our master's-level setting.

\balance
\section{Conclusion}
Three years of classroom use changed performance analysis from a last-week add-on into first-day practice. 
In the original setting, operational difficulty weakened the perceived relevance of the topic: some students considered it unnecessary, and only a minority independently found optimization potential with TAU. 
With EduMPI Suite embedded throughout MPI practice, performance evidence became expected feedback. 
Students used it to reason about program behavior, motivate improvements, and present their results, and all 33 participants in the latest evaluation recommended the tool for education.

The two historical cohorts document the original tension between interest, perceived importance, and independent use. 
Students valued guided work with TAU and wanted analysis earlier, yet few independently identified optimization potential and several reported substantial barriers.
The 2025 item-level comparison shows a uniform advantage for the representational functions needed in the scaffold. 
The feedback-to-transfer pattern explains how this change can be reproduced without making EduMPI Suite the final destination: immediate evidence, a visual communication model, explicit collective algorithms, and professional tools once learners can formulate questions. 
Future studies should quantify the effect more precisely and test delayed transfer, but the present longitudinal evidence supports moving performance analysis earlier and making it recurring MPI practice.

\section*{Acknowledgment}
The authors used AI assistants to help polish the wording, grammar, and clarity of the manuscript. All technical content, including the research ideas, claims, methodology, data analysis, and results, was produced and verified by the authors.

\printbibliography

\end{document}